\documentclass[traditabstract]{aa}
  
\usepackage{amsmath}
\usepackage{graphicx}
\usepackage{natbib}
\usepackage{txfonts}
\usepackage{indentfirst}

\newcommand{\eq}[1]{\begin{equation}  #1 \end{equation}}
\newcommand{\eqs}[1]{\begin{equation} \begin{split} #1 \end{split} \end{equation}}

\newcommand{\br}[1]{\left( #1 \right)}

\newcommand{\bb}[1]{\left[ #1 \right]}
\newcommand{\ba}[1]{\left\langle #1 \right\rangle}

\newcommand{\dd}{{\rm d}}
\newcommand{\expo}[1]{~{\rm e}^{ #1 }}
\newcommand{\vek}[1]{\mbox{\boldmath $#1$}}
\newcommand{\ic}{{\rm i}}

\newcommand{\rund}[1]{\left(#1\right)}

\def\Re{{\mathcal R}\hbox{e}}
\def\Im{{\mathcal I}\hbox{m}}

\begin{document}

\title{How well do third-order aperture mass statistics separate E- and
B-modes?}

\author{X. Shi \inst{1} \and B. Joachimi \inst{2} \and P. Schneider \inst{3}}

\offprints{X. Shi,\\
   \email{xun@mpa-garching.mpg.de}
}
\institute{Max-Planck-Institut f\"ur Astrophysik, Karl-Schwarzschild-Stra{\ss}e 1,
85740 Garching bei M\"unchen, Germany
\and Institute for Astronomy, University of Edinburgh, Royal Observatory,
Blackford Hill, Edinburgh, EH9 3HJ, U.K. 
\and Argelander-Institut f\"ur Astronomie (AIfA), Universit\"at Bonn, Auf
dem H\"ugel 71, 53121 Bonn, Germany}

\date{Received  / Accepted }

\abstract{With third-order statistics of gravitational shear it will be
possible to extract valuable cosmological information from ongoing and future
weak lensing surveys that is not contained in standard second-order statistics
because of the non-Gaussianity of the shear field. Aperture mass statistics are
an appropriate choice for third-order statistics because of their simple form
and their ability to separate E- and B-modes of the shear.
However, it has been demonstrated that second-order aperture mass statistics suffer
from E-/B-mode mixing because it is impossible to reliably estimate the shapes
of close pairs of galaxies. This finding has triggered developments of several
new second-order statistical measures for cosmic shear.  Whether the same
developments are needed for third-order shear statistics is largely determined by
how severe this E-/B-mixing is for third-order statistics. We tested third-order
aperture mass statistics against E-/B-mode mixing and found that the level of
contamination is well described by a function of $\theta/\theta_{\rm min}$,
where $\theta_{\rm min}$ is the cutoff scale. At angular scales of $\theta > 10
\;\theta_{\rm min}$, the decrease in the E-mode signal due to E-/B-mode
mixing is lower than 1 percent, and the leakage into B-modes is even less. For typical
small-scale cutoffs this E-/B-mixing is negligible on scales larger than a few arcminutes.
Therefore, third-order aperture mass statistics can safely be used to separate E-
and B-modes and infer cosmological information, for ground-based surveys as
well as forthcoming space-based surveys such as Euclid.
  } 

\maketitle

\section{Introduction}
Forthcoming large-field multi-colour imaging surveys, such as
KiDS\footnote{\texttt{http://www.astro-wise.org/projects/KIDS}},
DES\footnote{\texttt{http://www.darkenergysurvey.org}},
HSC\footnote{\texttt{http://www.naoj.org/Projects/HSC/index.html}},
LSST\footnote{\texttt{http://www.lsst.org}}, and
\textit{Euclid}\footnote{\texttt{http://www.euclid-ec.org}; \citet{laureijs11}},
will obtain galaxy shape and photometric redshift information for a huge number
of galaxies. This will boost weak lensing statistical power, especially in
constraining the properties of dark matter, dark energy, and the laws of
gravity.

The cosmological information obtained from cosmological weak lensing
can be enhanced by going beyond the standard analysis of second-order
(two-point) statistics. The most straightforward path to the
exploitation of higher-order statistical information is the use of
three-point functions of gravitational shear. These can probe
non-Gaussian signatures in the underlying matter density field, and
thus are key tools for better exploiting the wealth of information on
small, nonlinear scales.  Moreover, adding third-order statistics to
the weak lensing analysis may substantially improve the
self-calibration of systematic effects \citep{hut06}.

Previous studies \citep[e.g.][]{TJ04} suggested that the strength of
the constraints on cosmological parameters from third-order weak lensing 
statistics alone are comparable to those from two-point statistics. Recently,
 \citet{kayo13} and \citet{kayo13b} estimated the cosmological information
from combined two- and three-point statistics taking into account
non-Gaussian error covariances as well as the cross-covariance between
the power spectrum and the bispectrum. They found that adding the
third-order information improves the dark energy figure-of-merit of
weak lensing two-point statistics alone by about 60$\,\%$. This
potential benefit comes without the need for additional observations,
so that an efficient extraction of weak lensing three-point
information is desirable.

The increasingly precise measurements of future weak lensing
observations call for synchronously improving measurement
accuracy. The major sources of weak lensing systematics lie in the
measurement process, specifically, in the galaxy shape measurement
\citep[e.g.][]{kitching12} and the determination of photometric
redshifts \citep[e.g.][]{abdalla08}. Additionally, there are
systematics originating from astrophysical processes, the most
worrisome being the intrinsic alignment of galaxy shapes (see
\citealp{sembo08,shi10} for work at the three-point level).  How much
weak lensing three-point statistics are affected by these systematic
effects is still uncertain to a large degree.

In this situation, sensitive and reliable systematics tests are of utmost
importance.  One such test is the decomposition of statistics of the
gravitational shear into electric field-like E-mode components and
magnetic field-like B-mode components \citep{crittenden02,s02}. The
cosmological weak lensing signal only generates E-modes to first
order, with the B-mode signal making less than a per-mil level
contribution for two-point statistics \citep{hilbert09}.  Hence, a
significant B-mode signal serves as a smoking gun for the presence of
systematics in the data, also at the three-point level.

For second-order statistics, several E-/B-mode separating statistical
measures in configuration space have been developed, all of which can
be obtained via a linear transformation of the two-point shear
correlation functions. The aperture mass statistics \citep{s98} are
conceptually, and in practice, the easiest to apply, but require a
measurement of the shear correlation function down to lag zero for
perfect separation into E- and B-modes. However, the correlation
functions are not measurable at small separation, for instance because of 
the overlap of galaxy images \citep{vwaer00}. Note that it is
impractical to extract aperture mass statistics directly from data because of
gaps and masked areas in the images.

\citet{kilbinger06} demonstrated that the lower limit on the angular
galaxy separation available for correlation function measurements
leads to a significant leakage of E-modes into the B-mode aperture
mass statistic on small scales. This E-/B-mode mixing reduces the
effectiveness of the B-mode signal as a channel for detecting
systematics and, if unaccounted for, causes biases in any cosmological
analyses performed with the E-mode signal.

To eliminate this undesirable E-/B-mode mixing, more sophisticated
two-point statistics have been developed, including the ring
statistics and the the complete orthogonal sets of EB mode
integrals \citep{SK07, eifler10, fu10, s10, asgari12, kilbinger13}. The
corresponding weight functions for these statistics used in the transformation from correlation functions
have support only on finite intervals $[\theta_{\rm min};\theta_{\rm
  max}]$ where $\theta_{\rm min}>0$, which implies that the shear
two-point correlation functions (2PCFs) are not required for
$\theta<\theta_{\rm min}$ to calculate these statistics, for which
therefore no E-/B-mode mixing results from a cutoff in the 2PCFs.

The statistics of choice at the three-point level for the direct
application to data are again the shear correlation functions, since
they are most straightforward to measure from the data (since they are
unaffected by holes and gaps in the data field). However, matters are
complicated by the existence of $2^3=8$ components of the shear
three-point correlation functions (3PCFs), and three arguments instead
of one, as well as some freedom in the choice of reference frame for
the triplet of angular positions \citep{s03a}. Unsurprisingly, it is
much more difficult to construct three-point statistics that can be
accurately decomposed into E- and B-modes (see \citealp{shi11} for a
derivation of the general conditions).

Nonetheless, third-order ring statistics have been developed that
allow for a clean E-/B-mode separation \citep{krause12}, albeit at the
price of processing data with a complicated filter function whose
practicability still needs to be demonstrated. In contrast,
third-order aperture statistics, which were the first E-/B-mode
separating statistics generalized to the three-point level
\citep{jarvis04,s05}, are relatively easy to compute theoretically and
to derive from data. Consequently, they have been predominantly
employed in observational analyses, from early detections
\citep[e.g.][]{pen03} to recent results from the
COSMOS\footnote{\texttt{http://cosmos.astro.caltech.edu}} \citep{semboloni11}
and CFHTLenS\footnote{\texttt{http://www.cfhtlens.org}} (Kilbinger et
al., in prep.) surveys.

Three-point aperture statistics are susceptible to E-/B-mode mixing
caused by the unavailability of shear 3PCF measurements at small
angular separation, just like their two-point counterparts. To test
whether they remain viable as simple and well-established cosmological
probes for forthcoming weak lensing analyses, we determine
analytically the amount of E-/B-mode leakage expected for typical
lensing surveys.

\section{Separating E- and B-modes in the cosmic shear signal}

\subsection{General E-/B-mode separation}
Mathematically speaking, E- and  B-modes are general decompositions of a
spin-2 polarization field based on parity symmetry, just as curl-free and
divergence-free components are those of a spin-1 vector field. Whereas the
E-mode can be derived from a scalar potential, the B-mode can be derived from a
pseudo-scalar potential \citep{stebbins96, kamion97, zal97,hu97, kamion98,
crittenden02, s02}.

We follow the notation of \citet{s02} and describe the E- and B-mode components
of the cosmic shear field by defining the complex lensing potential
\eq{
\label{eq:psidef}
\psi \equiv \psi_{\rm E} + \ic \psi_{\rm B} \,.
}
Then the Cartesian components of the shear can be defined as
\eq{
\label{eq:defgamma}
\gamma \equiv \gamma_1 + \ic \gamma_2 = \frac{1}{2}
\br{\frac{\partial}{\partial x} + \ic \frac{\partial}{\partial y}}^2 \psi  \,.
}
Shear components defined in this way are not invariant under
coordinate rotation. One common way to amend this is to define the
tangential (t) and cross ($\times$) components of the shear
relative to a reference point on the two-dimensional plane
\citep{kamion98,crittenden02,s02}, 
\eq{
\label{eq:gcen}
\gamma_{\rm cen} \equiv \gamma_{\rm t} + \ic \gamma_{\times} = - \gamma
\expo{-2\ic\phi_{\vec{r}}}  \,,
}
where $\phi_{r}$ is the polar angle of the position vector connecting
the reference point to the point where the shear is measured. 
For second- and third-order statistics, we choose the reference point to be the
center of mass throughout this paper.

The goal of cosmic shear E-/B-mode separation is to find statistics
that respond only to the E-mode shear component of the field, and
B-mode statistics that are affected solely by the B-mode signal.
Taking correlation functions of the shear field as the `observables',
the general method to construct E-/B-mode statistics is to weight the
correlation functions, and find the conditions that these weight
functions need to satisfy to separate E- and B-modes.

At the two-point level, the `observables' of the shear field are
the shear 2PCFs
\eq{
\xi_+(r) \equiv \ba{\gamma_{\rm
cen}\gamma_{\rm cen}^*}(r) = \ba{\gamma \gamma^*}(r)\,,
} 
and 
\eq{
\xi_-(r)
\equiv \ba{\gamma_{\rm cen} \gamma_{\rm cen}}(r) = \langle \gamma \gamma
\rangle(\vek{r}) \expo{-4\ic \phi_r} \,,
}
for an angular separation $r$, where $\phi_r$ is the polar angle of
the vector $\vek r$ connecting the two points.  The imaginary
component of $\xi_-$ is expected to vanish because of parity invariance.

The general second-order E- and B-mode statistics can be defined as
\citep{SK07} \eqs{
\label{eq:EEandBB}
\textrm{EE} &= \int_0^{\infty} \vartheta\; \dd \vartheta\;\bb{ \xi_+(\vartheta)
T_+(\vartheta) + \xi_-(\vartheta) T_-(\vartheta) }\,,\\
\textrm{BB} &= \int_0^{\infty} \vartheta\; \dd \vartheta\;\bb{ \xi_+(\vartheta)
T_+(\vartheta) - \xi_-(\vartheta) T_-(\vartheta) }\,,
}
Under the condition that  
\eqs{
\label{eq:2pEBcondition}
\int_0^{\infty} & \vartheta\; \dd \vartheta\; T_+(\vartheta) J_0(\ell \vartheta)
= \int_0^{\infty} \vartheta\; \dd \vartheta\; T_-(\vartheta) J_4(\ell
\vartheta)\,, \\
& \rm{or\ equivalently}\\
&T_+(\vartheta) = T_-(\vartheta) + \int_{\vartheta}^{\infty} \theta\; \dd
\theta\; T_-(\theta) \br{\frac{4}{\theta^2} - \frac{12\vartheta^2}{\theta^4}}\,, }
EE only contains E-modes, BB only B-modes. One can also define a mixed-term
$\textrm{EB}$ which is derivable from a mixture of E- and B-modes
$\ba{\psi_{\rm E}\psi_{\rm B}}$, and observable as the imaginary part of
$\xi_-$. However, this EB term violates parity. It generally vanishes
since the shear field is expected to be parity symmetric.

E-/B-mode separation needs to be performed separately at each statistical order. 
The general approach to constructing E-/B-separating statistics remains the same
for higher-order statistics, only the conditions for the weight functions that
connect the shear correlation functions to the E-/B-separating statistics need
to be individually derived at each order.  For third-order statistics, the
conditions for general E-/B-mode separation are given in
\citet{shi11}.

\subsection{Aperture mass statistics}

The aperture mass $M_{\rm ap}$ was first introduced by
\citet{kaiser95} and \citet{s96} to estimate masses of galaxy
clusters from gravitational lensing signals. It is a
filtered version of both the shear $\gamma$ and the real part of the
convergence, $\kappa_E \equiv \nabla^2 \psi_E/2$ with  axisymmetric
filter functions,
\eqs{
\label{eq:mapdef}
M_{\rm ap}(\theta) &= \int \dd^2 r \; Q_{\theta}(|\vek{r}|)\; \gamma_{\rm t}(\vek{r}) \\
&=\int \dd^2 r \; U_{\theta}(|\vek{r}|)\; \kappa_E(\vek{r}) \,,
}
with tangential shear $\gamma_{\rm t}$ being specified relative to the
center of the aperture. The function $U_{\theta}$ is an compensated filter
function, and the filter functions $U_{\theta}$ and $Q_{\theta}$ are
inter-related by the relation \eq{
Q_{\theta}(\vartheta) = \frac{2}{\vartheta^2}\int_0^{\vartheta} \dd \vartheta'\;\vartheta' U_{\theta}(\vartheta') - U(\vartheta)\,.
}
The two most often used sets of filter function forms, the polynomial one
proposed by \citet{s98}, and the exponential one by \citet{crittenden02}, have (nearly) finite support in both real
and Fourier space. This feature makes the variance of aperture mass
$\ba{M_{\rm ap}^2}$ useful also in cosmic shear studies.  This
statistic provides a well-localized probe of the power spectrum and
is easy to determine from observational data.

Tangential shear averaged over a circle is only sensitive to E-modes,
whereas cross shear over a circle is only sensitive to B-modes. Therefore the
aperture mass $M_{\rm ap}$ is a measure of the E-mode shear. A corresponding
quantity,
\eq{
M_{\perp}(\theta) = \int \dd^2 r \; Q_{\theta}(|\vek{r}|)\; \gamma_{\times}(\vek{r})\,, 
}
accordingly is a measure of the B-mode shear.

\begin{figure}[t]
\centering
\includegraphics[width=8cm]{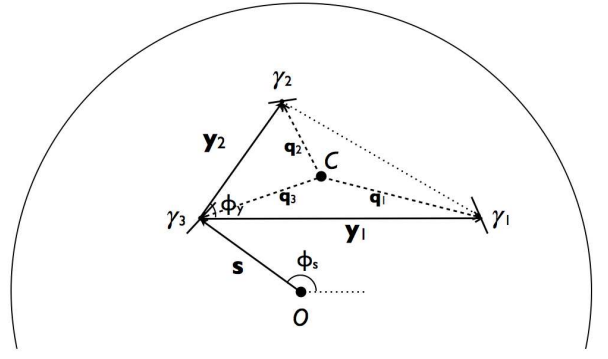}
\caption{Sketch of the halo model (1-halo term) for shear three-point functions
and the notations used. In the
text, the three shears in three-point shear correlator
e.g. $\ba{\gamma\gamma\gamma}$ correspond to  
$\gamma_1$, $\gamma_2$, and $\gamma_3$, accordingly. Note that the polar angle
of any vector $\vek{k}$ in Cartesian coordinates is denoted as $\phi_k$, but
$\phi_y$ is defined to be the angle between $\vek{y_1}$ and $\vek{y_2}$.}
\label{fig:0}
\end{figure}

At the three-point level, it is convenient to combine the eight correlation
functions into the natural components \citep{s03a} 
\eqs{
\label{eq:natural}
 \tilde{\Gamma}_{\rm cen}^{(0)}&(y_1,y_2,\phi_y) := \ba{\gamma_{\rm
cen}(\vek{s})\gamma_{\rm cen}(\vek{s}+\vek{y_1})\gamma_{\rm
cen}(\vek{s}+\vek{y_2})}
\\
=&
\ba{\gamma_{\rm t}\gamma_{\rm t}\gamma_{\rm t}} - \ba{\gamma_{\rm t}\gamma_{\times}\gamma_{\times}} -
\ba{\gamma_{\times}\gamma_{\rm t}\gamma_{\times}} -
\ba{\gamma_{\times}\gamma_{\times}\gamma_{\rm t}} + \\
& \ic\bb{\ba{\gamma_{\rm t}\gamma_{\rm t}\gamma_{\times}}+
\ba{\gamma_{\rm t}\gamma_{\times}\gamma_{\rm t}}+
\ba{\gamma_{\times}\gamma_{\rm t}\gamma_{\rm t}}-
\ba{\gamma_{\times}\gamma_{\times}\gamma_{\times}} } \,,\\
 \tilde{\Gamma}_{\rm cen}^{(3)}&(y_1,y_2,\phi_y) := \ba{\gamma_{\rm
cen}(\vek{s})\gamma_{\rm cen}(\vek{s}+\vek{y_1})\gamma^*_{\rm
cen}(\vek{s}+\vek{y_2})} \\
=&
\ba{\gamma_{\rm t}\gamma_{\rm t}\gamma_{\rm t}} + \ba{\gamma_{\rm t}\gamma_{\times}\gamma_{\times}} +
\ba{\gamma_{\times}\gamma_{\rm t}\gamma_{\times}} -
\ba{\gamma_{\times}\gamma_{\times}\gamma_{\rm t}} + \\
& \ic\bb{-\ba{\gamma_{\rm t}\gamma_{\rm t}\gamma_{\times}}+
\ba{\gamma_{\rm t}\gamma_{\times}\gamma_{\rm t}}+
\ba{\gamma_{\times}\gamma_{\rm t}\gamma_{\rm t}}+
\ba{\gamma_{\times}\gamma_{\times}\gamma_{\times}} } \,,
} where $\gamma_{\rm cen}$ is defined relative to the center of mass of the
triangle characterized with two side lengths $y_1, y_2$ and the angle between
them $\phi_y$ (see Fig.\;\ref{fig:0}). 
Because they are measured relative to a center of the triangle
formed by the three positions of shear measurements, these natural components are independent
of the orientation of the triangle. 

\citet{jarvis04} and \citet{s05} have derived the relations
between the natural components of the shear three-point functions
(shear 3PCFs) and the aperture mass statistics using the \citet{crittenden02}
filter functions
\eq{
U_{\theta}(\vartheta) = \frac{1}{2\pi\theta^2}
\br{1-\frac{\vartheta^2}{2\theta^2}} \expo{-\vartheta^2/2\theta^2} \,,
}
\eq{
\label{eq:Qform}
Q_{\theta}(\vartheta) = \frac{\vartheta^2}{4\pi\theta^4}
\expo{-\vartheta^2/2\theta^2} \,.
}
Here we present the relations in the specific case of three equal filter sizes, and refer
to equations (62), (68) and (73) in \citet{s05} for the general form:
\eqs{
\label{eq:map_3}
\ba{M_{\rm ap}^3}(\theta) \;=\;&  \frac{1}{4}\int_0^{\infty} \frac{\dd y_1
y_1}{\theta^2}  \int_0^{\infty} \frac{\dd y_2 y_2}{\theta^2}  \int_0^{2\pi}
\frac{\dd \phi_y}{2\pi}\,   \\
&  \Re\; \Bigl[ 3 \tilde{\Gamma}_{\rm cen}^{(3)}(y_1,y_2,\phi_y)
T_3(\theta,y_1,y_2,\phi_y) \\
& + \tilde{\Gamma}_{\rm
cen}^{(0)}(y_1,y_2,\phi_y) T_0(\theta,y_1,y_2,\phi_y) \Bigr]
}
is the pure E-mode (EEE) statistics, with

\eq{
\label{eq:T0}
T_0(\theta,y_1,y_2,\phi_y) = \frac{1}{24} \frac{q_1^2 q_2^2 q_3^2}{\theta^6}
\exp{\rund{-\frac{q_1^2+q_2^2+q_3^2}{2\theta^2}}}
}
and
\eqs{
\label{eq:T3}
&T_3(\theta,y_1,y_2,\phi_y) = \exp{\rund{-\frac{q_1^2+q_2^2+q_3^2}{2\theta^2}}}
\;\\
& \times \bb{\frac{1}{24}\frac{q_1^2 q_2^2 q_3^2}{\theta^6} -
\frac{1}{9}\frac{\vek{q_1}\vek{q_2}\vek{q_3}^{*2}}{\theta^4}
+\frac{1}{27}\br{\frac{\vek{q_1}^2\vek{q_2}^2\vek{q_3}^{*4}}{q_1^2 q_2^2 q_3^2
\theta^2} + \frac{2\vek{q_1}\vek{q_2}\vek{q_3}^{*2}}{q_3^2\theta^2} } } \;,
}
where $\vek{q}_i$ indicates the vectors pointing from the center of mass of
the triangle $(y_1,y_2,\phi_y)$ to its
vertices (see Fig.\;\ref{fig:0}),  
\eqs{
\vek{q_1} = \frac{2\vek{y_1}-\vek{y_2}}{3} \,; \vek{q_2} =
\frac{2\vek{y_2}-\vek{y_1}}{3} \,; \vek{q_3} = -\frac{\vek{y_1}+\vek{y_2}}{3}
\;. } 
Here we used a complex notation so that a vector $(a,b)$ corresponds to the
complex number $\vek{q} = a + \ic b$. The asterisk denotes the complex
conjugation, so $\vek{q}^*=a - \ic b$ corresponds to the
vector $(a,-b)$. For the
forms of the weight functions $T_{0,3}$, a Gaussian filter was
assumed. Without loss of generality, we choose $\vek{y_1}$ to be 
parallel to the $x$-axis, so that $\phi_{y_1}=0$ and $\phi_y$ is the polar angle
of $\vek{y}_2$.

Similarly, the B-mode three-point aperture mass statistics include the EEB term
\eqs{
 \ba{M_{\rm ap}^2 M_{\perp}}(\theta) \;=\;&  \frac{1}{4}\int_0^{\infty}
 \frac{\dd y_1 y_1}{\theta^2}  \int_0^{\infty} \frac{\dd y_2 y_2}{\theta^2}  \int_0^{2\pi}
\frac{\dd \phi_y}{2\pi}\,   \\
&\Im\; \Bigl[ \tilde{\Gamma}_{\rm cen}^{(3)}(y_1,y_2,\phi_y)
T_3(\theta,y_1,y_2,\phi_y) \\
& +\tilde{\Gamma}_{\rm cen}^{(0)}(y_1,y_2,\phi_y)
T_0(\theta,y_1,y_2,\phi_y)  \Bigr]  \;,
}
the EBB term
\eqs{
\label{eq:EBB}
 \ba{M_{\rm ap} M_{\perp}^2}(\theta)\; =\; & \frac{1}{4}\int_0^{\infty}
 \frac{\dd y_1 y_1}{\theta^2}  \int_0^{\infty} \frac{\dd y_2 y_2}{\theta^2}  \int_0^{2\pi}
\frac{\dd \phi_y}{2\pi}\,\\
& \Re \; \Bigl[ \tilde{\Gamma}_{\rm cen}^{(3)}(y_1,y_2,\phi_y)
T_3(\theta,y_1,y_2,\phi_y)\\
&-\tilde{\Gamma}_{\rm cen}^{(0)}(y_1,y_2,\phi_y)
T_0(\theta,y_1,y_2,\phi_y)\Bigr] \;,
}
and the BBB term
\eqs{
\label{eq:BBB}
 \ba{M_{\perp}^3}(\theta) \;=\;&  \frac{1}{4}\int_0^{\infty} \frac{\dd y_1
y_1}{\theta^2}  \int_0^{\infty} \frac{\dd y_2 y_2}{\theta^2}  \int_0^{2\pi}
\frac{\dd \phi_y}{2\pi} \\
& \Im \;\Bigl[ 3 \tilde{\Gamma}_{\rm cen}^{(3)}(y_1,y_2,\phi_y)
T_3(\theta,y_1,y_2,\phi_y) \\
&-\tilde{\Gamma}_{\rm cen}^{(0)}(y_1,y_2,\phi_y)
T_0(\theta,y_1,y_2,\phi_y)\Bigr]  \;.
}

Among them, $\ba{M_{\rm ap}^2 M_{\perp}}$ and $\ba{M_{\perp}^3}$ violate parity,
whereas $\ba{M_{\rm ap} M_{\perp}^2}$ conserves parity symmetry and therefore is harder to distinguish from pure E-mode statistics.

\section{E-/B-mode mixing with three-point aperture mass statistics}
Here we investigate the degree of EB-mixing in three-point aperture
mass statistics due to the aforementioned small-scale information
loss. Since three-point statistics probe more into the nonlinear,
small-scale regime than their two-point counterparts, one might
naively expect three-point statistics to be more severely affected by
the small-scale information loss.

We follow these steps: (i) we model the `observable' shear
3PCFs which contain only E-mode signal; (ii)
we derive the aperture statistics from the shear 3PCFs; (iii) we introduce a
cutoff in the 3PCFs at a small angular scale, mimicking the effect of information loss
due to unreliable shear estimates caused by overlapping galaxy images (this
generates a B-mode signal that results in a nonzero B-mode
aperture statistic $\ba{M_{\rm ap} M_{\perp}^2}$ that conserves parity), and
repeat step (ii) using only correlation functions above the cutoff scale; (iv)
finally, we compare the aperture mass statistics resulted from steps (ii) and
(iii).

\subsection{Modeling}

We use the real-space halo model \citep{zal03,TJ03c} to model shear
three-point correlation functions. Since our study focuses on
small scales, the halo model is a natural choice because it is expected
to provide the most precise predictions on small scales. For the same
reason, slightly imprecise modeling on large scales will not affect
our main results. We take advantage of this and neglect the two- and
three-halo terms, which significantly reduces the computational load. In
fact, as shown by \citet{TJ03c}, the one-halo term already captures most
of the features of the shear 3PCFs measured from ray-tracing
simulations.

In the real-space halo model the convergence profile for a halo of mass $M$ can
be expressed as
\eqs{
\label{eq:gammam0}
\kappa(s, M, \chi, \chi_{\rm s}) &\equiv  \frac{\Sigma(M,\chi,s)}{\Sigma_{\rm
crit}(\chi,\chi_s)}\\
&=\frac{4\pi G a(\chi)}{c^2 }
\frac{ f_K(\chi)f_K(\chi_{\rm s}-\chi)}{f_K(\chi_{\rm
s})} \Sigma(s, M,\chi) \,, }
where $s$ is the angular distance from the center of the halo, $f_K$ indicates
the comoving angular diameter distance, $a$ is the scale factor, and $\chi$ and
$\chi_s$ are the comoving distance to the halo and the source, respectively. The
surface mass density $\Sigma$ contributed by the halo is computed from an NFW
profile $\rho_{\rm NFW}$ \citep{nfw} with cutoff at the virial radius $r_{\rm
vir}$ \eq{ \Sigma(s, M,\chi) =  \int_{-\sqrt{r^2_{\rm vir}-a^2 f_K^2 s^2}}^{\sqrt{r^2_{\rm
vir}-a^2 f_K^2 s^2}} \dd
r_{\parallel}\; \rho_{\rm NFW}\br{M, \sqrt{r_{\parallel}^2 +
a^2(\chi)f_K^2(\chi) s^2}} \,, }
where $r_{\parallel}$ is the proper distance along the line of sight.
$\Sigma$ can be expressed analytically, 
\eq{
\Sigma(s, M,\chi) = \frac{M}{2\pi r_{\rm vir}^2} \frac{c_{\rm
h}^2\; F(c_{\rm h} s/\theta_{\rm vir})}{\ln(1+c_{\rm h}) - c_{\rm
h}/(1+c_{\rm h})}\,,
} 
with $c_{\rm h}$ being the halo concentration parameter, and $\theta_{\rm
vir}(M,\chi) = r_{\rm vir}(M, \chi)/ f_K(\chi)/a(\chi)$. The explicit
form of the function $F(x)$ is given by equation (27) in \citet{TJ03a}. 

The halo model tangential shear profile can then be expressed using the relation
\eq{
\label{eq:gammam}
\gamma_{\rm t}(s,M,\chi,\chi_{\rm s}) = \bar{\kappa}(s,M,\chi,\chi_{\rm s}) -
\kappa(s,M,\chi,\chi_{\rm s}) \,, }
with $\bar{\kappa}(s,M,\chi,\chi_{\rm s})$ being the mean surface mass density
inside a circle of radius $s$. The tangential shear profile can also be expressed analytically
as
\eqs{
\label{eq:gt}
\gamma_{\rm t}(s,M,\chi,\chi_{\rm s}) = &\frac{2 G M a(\chi)}{c^2 r_{\rm vir}^2}
\frac{f_K(\chi)f_K(\chi_{\rm s}-\chi)}{f_K(\chi_{\rm
s})} \\
 & \times \frac{c_{\rm
h}^2\; G(c_{\rm h} s/\theta_{\rm vir})}{\ln(1+c_{\rm h}) - c_{\rm
h}/(1+c_{\rm h})}\,,
}
see equations (17) in \citet{TJ03c} for the explicit form of the function
$G(x)$. Note that comoving coordinates have been used to express the surface
mass density and the distances in \citet{TJ03a,TJ03c}.

The natural
components of shear 3PCFs can then be derived by averaging shear triplets over
the triangle angular position $\vec{s}$, and summing contributions from all
halos with a comoving distance $\chi$ smaller than that of the source $\chi_s$ 
\citep[see also][]{TJ03a}, \eqs{
\label{eq:Gi}
& \tilde{\Gamma}_{\rm cen}^{(i)}(y_1,y_2,\phi_y) =
\Omega \int^{\chi_s}_{0}\!\!\dd \chi~ \frac{\dd^2 V}{\dd \chi \dd \Omega}
\int\!\!\dd M~ \frac{\dd n(M,\chi)}{\dd M} \\
& \int_{\Omega}\; \!\! \frac{\dd^2s}{\Omega} \;    
\gamma_{\rm t}(|\vec{s}+\vec{y_1}|) \gamma_{\rm
t}(|\vec{s}+\vec{y_2}|) \gamma_{\rm t}(s)\; P^{(i)}_{\rm
cen}(\vec{s},\vec{y_1},\vec{y_2}) \\ 
 = & \int^{\chi_s}_{0}\!\!\dd\chi~ f_K^2(\chi)
\int\!\!\dd M~ \frac{\dd n(M,\chi)}{\dd M} \int_{\Omega}\;\dd^2s \\
&    \; \gamma_{\rm t}(|\vec{s}+\vec{y_1}|) \gamma_{\rm
t}(|\vec{s}+\vec{y_2}|) \gamma_{\rm t}(s)\; P^{(i)}_{\rm
cen}(\vec{s},\vec{y_1},\vec{y_2}) \,,
}
where $\Omega$ is the angular coverage of the full sky, $V$ is the comoving
volume element, and $n$ is the comoving halo mass function.
We have taken the form of the comoving differential volume $\dd^2 V/\dd \chi \dd
\Omega = f_K^2$, which is true for a flat universe.
Again we chose $y_1$ to be parallel to the $x$-axis. This does not affect
generality since the last integral does not depend on $\phi_{y_1}$ because of
the spherical symmetry of a halo. The mass and distance dependencies of $\gamma_{\rm t}$ are not shown to keep the notation concise. We define $P^{(i)}_{\rm
cen}(\vec{s},\vec{y_1},\vec{y_2})$ to be the projection operator for
$\tilde{\Gamma}_{\rm cen}^{(i)}$, which projects the three shears (or their
complex conjugates) measured relative to the center of the halo to those measured relative to the center of mass of the triangle formed by $\vec{y_1}$ and $\vec{y_2}$ (see Fig.\;\ref{fig:0}). 
To derive its form, we first
consider the projection of one of the shears, for example, $\gamma_3$. The shear
value $\gamma(s)$ can be seen as $\gamma_3$ measured relative to the center of
the halo. Using Eq.\;(\ref{eq:gcen}), one can express $\gamma_3$ in Cartesian
coordinates and then project it relative to the center of mass of the triangle 
\eqs{
\gamma_{3,\rm{cen}}\; =\; & -\expo{-2\ic\phi_{\vec{q_3}}}
\bb{-\expo{2\ic\phi_{\vec{s}}}}\; \gamma_{\rm t}(s) \\
= \;&\expo{2\ic\br{\phi_{\vec{s}} - \phi_{\vec{q_3}}}}\; \gamma_{\rm t}(s) \,.
 } 
 Taking the definition of the natural components (\ref{eq:natural}) into
 account, the explicit form of $P^{(i)}_{\rm cen}$ reads 
 
 \eqs{ 
\ P^{(0)}_{\rm cen} &= \exp\bb{2\ic \br{
\phi_{\vec{s}+\vec{y_1}} -\phi_{\vec{q_1}} +
 \phi_{\vec{s}+\vec{y_2}}-\phi_{\vec{q_2}} +
 \phi_{\vec{s}} - \phi_{\vec{q_3}}}}  \,, \\
\ P^{(3)}_{\rm cen} &= \exp\bb{2\ic \br{
\phi_{\vec{s}+\vec{y_1}} -\phi_{\vec{q_1}} +
 \phi_{\vec{s}+\vec{y_2}}-\phi_{\vec{q_2}} -
 \phi_{\vec{s}} + \phi_{\vec{q_3}}}} \,. \\
}
The different polar angles can be expressed as functions of  
$\vec{s}$, $\vec{y_1}$ and $\vec{y_2}$.

We adopt a flat $\Lambda$CDM cosmology with
dark matter content $\Omega_{\rm m0}=0.28$, dark energy
content $\Omega_{\Lambda}=0.72$, Hubble parameter $h_0=0.7$, slope of
the initial power spectrum $n_s=0.96$, and the normalization of the matter power
spectrum $\sigma_8=0.8$. We use the \citet{tinker08} formula to model the
halo mass function, and the \citet{duffy08} fitting formula for the
concentration parameter of NFW halos.
To reduce the computational load even more, we assume a single plane of source
galaxies at redshift $z_{\rm s}=1$ (see the discussion section for a
generalization from this case).

\begin{figure}[h]
\centering
\includegraphics[width=7cm]{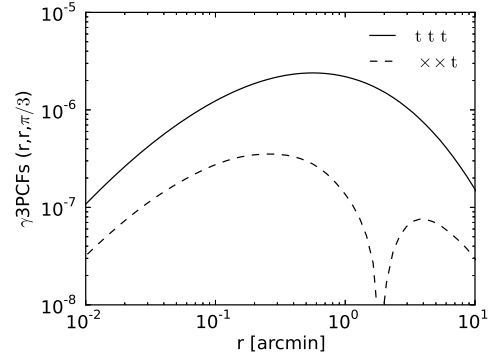}
\caption{Dependence of the modeled shear 3PCFs on angular separation. 
The two parity-invariant components ttt (stands for $\ba{\gamma_{\rm t}
\gamma_{\rm t} \gamma_{\rm t}}$) and $\times\times$t (stands for $\ba{\gamma_{\times} \gamma_{\times} \gamma_{\rm t}}$) are shown for equilateral
triangles, see equation (\ref{eq:natural}). The $\times\times$t signal becomes
negative beyond $r \approx 2$ arcminutes, and its absolute value is
plotted for larger angles.}
\label{fig:1}
\end{figure}

\begin{figure}[h]
\centering
\includegraphics[width=9cm]{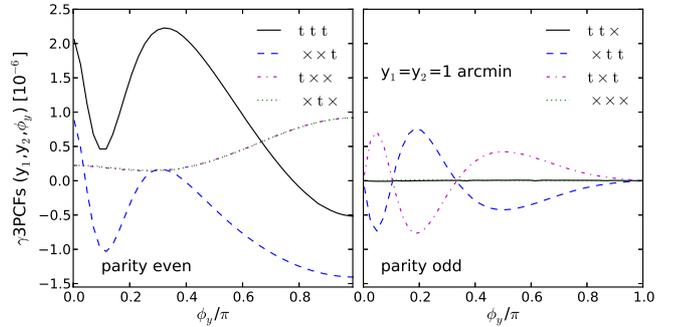}
\caption{Dependence of the modeled  shear 3PCFs on triangle configuration. A
subset of triangle configurations with two equal side-lengths $y_1 = y_2 = 1$
arcminute is shown, and plotted are the dependencies of the shear 3PCFs on the angle
between them. The left panel `parity even' shows shear 3PCFs that involve
zero or two tangential shears and are thus invariant under a parity
transformation, while the right panel shows shear 3PCFs that contain an odd
number of tangential shears and therefore change sign under parity transformation. The notation in the
labels is analogous to Fig.\;\ref{fig:1}.
}
\label{fig:2}
\end{figure}
 
Figures \ref{fig:1} and \ref{fig:2} present the modeled shear
three-point correlation functions. Shown are the three-point correlators of
shear components that can be seen as parts of the natural components, see
(\ref{eq:natural}). As can be seen from Fig.\;\ref{fig:1}, the pure tangential component
$\ba{\gamma_{\rm t}\gamma_{\rm t}\gamma_{\rm t}}$ peaks at about $30$ arcseconds
for equilateral triangle configurations. In comparison, the virial radius
for a $10^{14}$ M$_{\odot}$ halo at $z=0.5$ corresponds to an angular
size of approximately $100$ arcseconds. In contrast to shear 2PCFs,
all shear 3PCFs vanish when $r\to 0$. This is expected since a
statistically isotropic spin-2 field has vanishing skewness.
For equilateral triangles,
$\ba{\gamma_{\times}\gamma_{\times}\gamma_{\rm t}}=\ba{\gamma_{\times}\gamma_{\rm t}\gamma_{\times}}=\ba{\gamma_{\rm t}\gamma_{\times}\gamma_{\times}}$,
and all the components involving $1$ or $3$ $\gamma_{\times}$
vanish, owing to the additional symmetry of three equal side-lengths. Therefore
only two components are plotted in Fig.\;\ref{fig:1}.

Figure \ref{fig:2} shows a strong dependence of shear 3PCFs on the
triangle shape. After taking account of the different ordering of the
three shears and the different definitions of $\phi$, the shapes are
consistent with Fig.\;6 in \citet{zal03}. 
The detailed shapes of the curves are difficult to understand in
detail, but several features do appear as expected. First,
for equilateral triangles ($\phi_y=\pi/3$) one sees the additional
symmetry $\ba{\gamma_{\times}\gamma_{\times}\gamma_{\rm
t}}=\ba{\gamma_{\times}\gamma_{\rm t}\gamma_{\times}}=\ba{\gamma_{\rm
t}\gamma_{\times}\gamma_{\times}}$ and
$\ba{\gamma_{\times}\gamma_{\rm t}\gamma_{\rm t}}=\ba{\gamma_{\rm
t}\gamma_{\times}\gamma_{\rm t}}=\ba{\gamma_{\rm t}\gamma_{\rm
t}\gamma_{\times}}$. Second, because we plotted isosceles triangles, for
all opening angles $\ba{\gamma_{\times}\gamma_{\rm t}\gamma_{\times}}=\ba{\gamma_{\rm t}\gamma_{\times}\gamma_{\times}}$, $\ba{\gamma_{\times}\gamma_{\rm
t}\gamma_{\rm t}}=-\ba{\gamma_{\rm t}\gamma_{\times}\gamma_{\rm t}}$, and
$\ba{\gamma_{\rm t}\gamma_{\rm t}\gamma_{\times}} =
\ba{\gamma_{\times}\gamma_{\times}\gamma_{\times}} = 0$.
Third, for collapsed triangles ($\theta=0$), the $\ba{\gamma_{\rm t}\gamma_{\rm t}\gamma_{\rm t}}$ and $\ba{\gamma_{\times}\gamma_{\times}\gamma_{\rm t}}$
components have the same sign. This is expected since for collapsed
triangles two of the three shear components have the same value and thus their
product is always positive. The sign of the shear 3PCF then depends on the third
shear component, which is shared by $\ba{\gamma_{\rm t}\gamma_{\rm
t}\gamma_{\rm t}}$ and $\ba{\gamma_{\times}\gamma_{\times}\gamma_{\rm t}}$.
Judging from this feature, some
of the shear 3PCFs in Fig.\;10 in \citet{TJ03c} have incorrect signs.
Apart from that figure shows the same dependencies of the shear 3PCFs. Finally,
the triangle configuration with opening angle $\phi_y =\psi$ and that with $\phi_y =-\psi$ can be considered as mirror images of each
other, so are the triangle configuration with opening angle $\phi_y = \pi -
\psi$ and that with $\phi_y = \pi + \psi$.
This implies that $\gamma$3PCF$(y_1,y_2,\psi)=\gamma$3PCF$(y_1,y_2,-\psi)
=\gamma$3PCF$(y_1,y_2,2\pi-\psi)$ when $\gamma$3PCF is a parity-even shear
3PCF, and therefore the slopes of the parity-even shear 3PCFs at $\phi_y=0$ and
$\phi_y=\pi$ are zero; on the other hand, $\gamma$3PCF$(y_1,y_2,\psi)=-\gamma$3PCF$(y_1,y_2,-\psi) =-\gamma$3PCF$(y_1,y_2,2\pi-\psi)$ when $\gamma$3PCF is a parity-odd shear
3PCF, and the shear 3PCF values at $\phi=0$ and
$\phi=\pi$ for parity-odd shear 3PCFs are zero, as shown in the figure.

\begin{figure}[h]
\centering
\includegraphics[width=7cm]{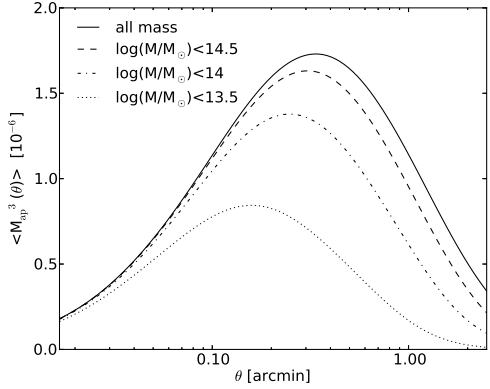}
\caption{Three-point aperture mass statistic $\langle M_{\rm
ap}^3 \rangle$ modeled with the halo model with the one-halo term. Contributions
from halos below different maximum mass limits are shown. }
\label{fig:3}
\end{figure}

\begin{figure}[h]
\centering
\includegraphics[width=7cm]{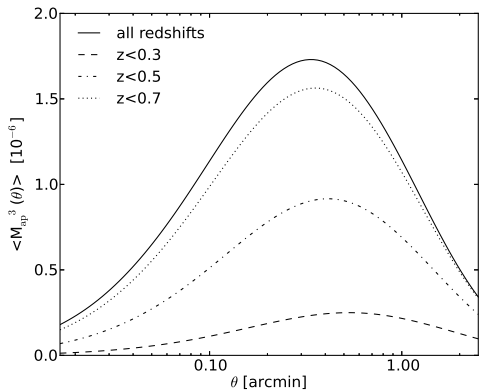}
\caption{Same as Fig.\;\ref{fig:3}, but showing contributions from halos
below different redshift limits. The line with 'all redshifts' contains
contributions from all halos below the source redshift $z_{\rm s} = 1$.}
\label{fig:32}
\end{figure}

The modeled third-order aperture mass statistic $\ba{M_{\rm ap}^3}$
under different mass and redshift cuts are presented in
Figs.\;\ref{fig:3} and \ref{fig:32}. Since only the one-halo term is
considered in the model, the contributions from different halos to the
$\ba{M_{\rm ap}^3}$ signal are additive.  The dominant contribution to
the signal on and above arcminute scales comes from $z=0.3$ to $z=0.7$, and from
massive halos corresponding to galaxy groups and clusters. In particular, when the filter scale is larger than $80$
arcseconds, more than half of the signal is contributed by halos with
masses larger than $10^{14}$ M$_{\odot}$. The peak of the signal is
located at $\theta \approx 20$ arcseconds.  The signal at larger aperture
radii is dominated by larger halos at lower redshifts.  Therefore the signal shifts to lower $\theta$ as an upper mass cutoff is introduced,
and to higher $\theta$ as an upper redshift cutoff is introduced.

\begin{figure}[h]
\centering
\includegraphics[width=9cm]{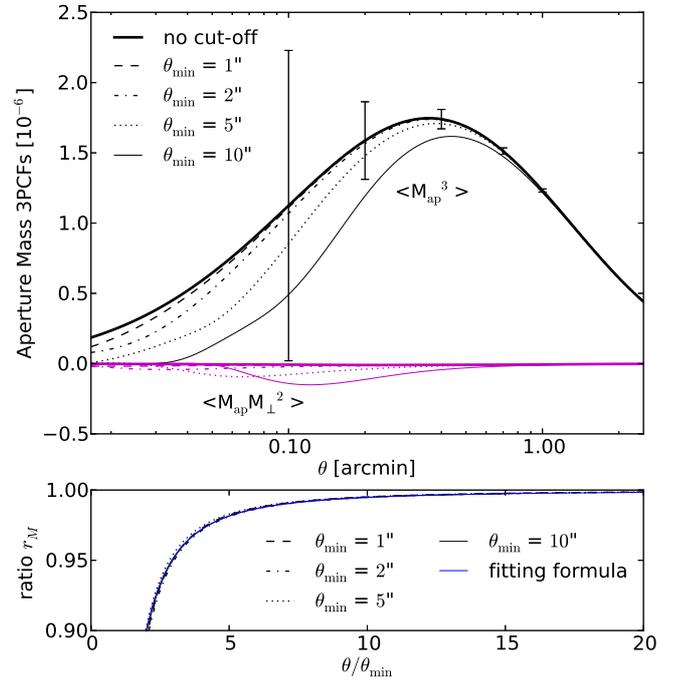}
\caption{E-/B-mode mixing with three-point aperture mass statistics due to
small-scale cutoff at $\theta_{\rm min} = $1, 2 and 5 arcseconds. 
In the upper
panel, the black lines represent the E-mode aperture mass statistic $\langle M_{\rm
ap}^3 \rangle$, and the magenta lines represent the parity-preserving B-mode
$\langle M_{\rm ap} M_{\perp}^2 \rangle$. 
The error bars present the shape noise contribution to the uncertainty
in $\langle M_{\rm ap}^3 \rangle$ at $\theta=0.1,0.2,0.4,0.7$, and $1$
arcmin. To compute the shape noise we have adopted an ellipticity dispersion
$\sigma_{\epsilon}$ = 0.35, a galaxy number density $n = 30$ arcmin$^{-2}$, and
a survey area of $15000$ deg$^2$, which are typical values for the Euclid
survey.
The lower panel presents the relative
decrease in $\langle M_{\rm ap}^3 \rangle$ signal when a small-scale information
loss is present. All ratios $r_M \equiv \langle M_{\rm ap}^3 \rangle(\theta,
\theta_{\rm min})/\langle M_{\rm ap}^3 \rangle(\theta,
\theta_{\rm min}=0)$ follow the same curve when plotted against
$\theta/\theta_{\rm min}$. The thin blue line presents a fitting formula of this
curve $r_M=1-1/(0.8+0.2x+2.3x^2-0.04 x^3)$ with $x=\theta/\theta_{\rm min}$. The
parity-violating B-mode signals, which are not explicitly shown in the figure,
are zero over all angular scales, both with and without cutoff. }
\label{fig:4}
\end{figure}

\subsection{Effect of small-scale cutoff}

As described in \citet{kilbinger06}, the small-scale information loss
is attributed to the fact that the shapes of close galaxy pairs cannot
be estimated reliably (see also the detailed discussion in
\citealp{miller13}). The angular scale below which this information loss
occurs depends on the true angular sizes of galaxies, on the point
spread function of the observation, and the ability of the shear
measurement algorithm to separate overlapping isophotes.  The
typical size of this cutoff used in current shear measurement methods
is a few arcseconds for space-based observations (e.g. 3 arcseconds in
the COSMOS analysis of \citealp{schrabback10}), and slightly larger
scales for ground-based observations (e.g. 9 arcseconds in CFHTLenS as
discussed in \citealp{kilbinger13}).

We introduce a small angular scale cutoff in the three-point
correlation function at 1, 2, 5, and 10 arcseconds, and show the
E-/B-mode mixing introduced into the aperture mass statistics $\langle
M_{\rm ap}^3 \rangle$ and $\langle M_{\rm ap} M_{\perp}^2 \rangle$ in
Fig.\;\ref{fig:4}.

As expected, small-scale information loss leads to a decrease in the
E-mode aperture mass signal $\ba{M_{\rm ap}^3}$, and introduces a
spurious parity-conserving B-mode signal $\langle M_{\rm ap}
M_{\perp}^2 \rangle$. The parity-violating B-mode signals are
consistent with zero over all angular scales, both with and without
cutoff.  This is expected since aperture mass statistics are computed
over shear correlation functions with triangle configurations of both
parities.

%why not strong compared to 2pt?
Somewhat surprisingly, the decrease in the E-mode aperture mass signal
due to a small-scale cutoff is less significant for third-order
statistics than at second order \citep[see Fig.\;1
of][]{kilbinger06}.  The angular scales that are significantly affected are
below $10\;\theta_{\rm min}$. This means that even for a conservative
small-scale cutoff at $10$ arcseconds, the effect is limited to $\lesssim$
arcminute scales. Since these scales are strongly influenced by complicated baryonic physics which prevents precise theoretical predictions,
signals on these small scales are usually not used to infer cosmology. 
At second-order,
however, the decrease at $1$ arcminute angular separation can be as
high as $10\%$ for $\theta_{\rm min}= 5$ arcseconds, compared to $<0.5\%$ at the
third-order with the same cutoff scale.
This difference can be understood by comparing the relations between aperture mass
statistics and shear correlations function for the second- and third-order
case. In the former case, one of the two correlation
functions $\xi_+$ and its corresponding weight function $T_+$ both
peak at zero angular separation. In contrast, none of the shear 3PCFs
peak at small scales, and both weight functions $T_0$ and $T_3$ vanish
when $\theta$ approaches $0$. Accordingly, a small-scale cutoff in the
three-point shear correlation functions causes a weaker effect on
the aperture mass statistics.

Leakage to B-modes is even weaker than the decrease in E-mode
signal. This is also very different from the two-point case, where the
growth in the B-mode signal on small scales due to a cutoff approximately
equals the decrease in E-mode signal in size. This small leakage
enables the parity conserving B-mode $\langle M_{\rm ap}
M_{\perp}^2 \rangle$ to remain a good test for systematics in third-order statistics even in the presence of
a small-scale cutoff in the 3PCFs.

At the small angular scales where the E-/B-mode mixing occurs, shot or
shape noise (the noise due to randomness of the intrinsic shapes of the
galaxies) is the major source of uncertainty in the measurement of the aperture
mass signals. We plot in Fig.\;\ref{fig:4} the shape noise contribution (see
appendix) to the uncertainty of $\ba{M_{\rm ap}^3}$
for the Euclid survey. The decrease in the
E-mode aperture mass signal caused by a small-scale cutoff is 
subdominant to the uncertainty of $\ba{M_{\rm ap}^3}$ measurement for
$\theta_{\rm min} < 5$ arcseconds, even for a large and relatively deep survey
like Euclid. For surveys with smaller sky coverage and/or shallower depth, the
shot noise contribution is expected to be even higher.
Therefore, the E-/B-mode mixing effect on the third-order statistics is
generally not observable.

\section{Discussion}
As mentioned above, a single source-plane at $z=1$ was used for this
study. This source redshift was chosen to represent deep surveys like Euclid
(which has an expected median source redshift of $z \approx 0.9$). The results we obtained, however, can be
generalized to shallower surveys as well. According to the halo model, if the comoving distance to the
source $\chi_s$ is changed, the shear 3PCF contributed by a
certain halo at $\chi$ changes its amplitude as $\bb{f_K(\chi_{\rm s} -
\chi)/f_K(\chi_{\rm s})}^3$, whereas the angular dependence is retained
(\ref{eq:gt}, \ref{eq:Gi}). A shallower survey will give more relative weight to low-redshift halos, which will shift the signals to larger angular scales (see e.g.  Fig.\;\ref{fig:32}), making $\ba{M_{\rm ap}^3}$ slightly less affected by small-scale cutoff.

We have tested EB-mixing in third-order aperture statistics with three
equal-sized filters. For the more general case with three different
filter sizes, that is $\ba{M_{\rm ap}^3}(\theta_1,\theta_2,\theta_3)$, we
argued that the degree of EB-mixing is bounded from above by that for
$\ba{M_{\rm ap}^3}(\theta_1)$, with $\theta_1$ being the smallest of
the three filter sizes. To see this, it is helpful to consider the
aperture mass as obtained by placing apertures on the image. In this
case, the estimator of $\ba{M_{\rm ap}^3}$ involves summation over all
galaxy triplets in the field, with the three galaxies in a triplet
being weighted by three different filters. A small-scale cutoff will
eliminate close triplets from this sum and cause the E-mode signal to
be underestimated. This underestimation is most severe when all three
galaxies in the triplet are given high weights from their corresponding
filters. Because the filter $Q_{\theta}$ is more extended for a larger filter
size $\theta$, the eliminated close triplets are given on average a lower
weight relative to the retained ones when some of the filter sizes are
larger in size. Therefore, if EB-mixing is negligible in $\ba{M_{\rm
    ap}^3}$ at $\theta_1$, it is also negligible in a general aperture
mass statistic $\ba{M_{\rm ap}^3}(\theta_1,\theta_2,\theta_3)$ with
$\theta_2 \geq \theta_1$ and $\theta_3 \geq \theta_1$.

The test we performed shows how much E-mode signal will leak to the B-mode
under a small-scale cutoff for third-order aperture statistics.  The
opposite question needs to be addressed in a quantitative way as well: if
there exists a B-mode signal, for instance, from noise and bias coming from
shape measurements or intrinsic alignments of galaxy shapes, how much
will the E-mode signal be affected?  The answer to this question
depends on the actual angular and shape dependencies of the particular
B-mode contamination and is therefore hard to give in general.  For
future lensing surveys, once the amplitude of the B-mode signal and its
statistical uncertainty is known, one can study the maximum bias a
small-scale cutoff could contribute to the E-mode signal. Ultimately,
each contribution to both E- and B-mode signals needs to be understood
to derive precise cosmological information from cosmic shear
measurements.

\section{Conclusion}

We tested the degree of EB-mixing in third-order aperture statistics due
to the inevitable absence of small-scale correlation
measurements. Both the decrease in E-mode signal and the introduction
of a spurious B-mode signal were found to be smaller for the third-order
aperture mass statistics than for the second-order ones. Quantitatively,
the change in the E-mode signal is lower than $1\%$ at an angular
separation of ten times the cutoff scale and above, and therefore is
negligible on angular scales of interest to ongoing and future weak lensing
surveys. Some parity-preserving B-mode signal is created on small
angular scales ($< 1$ arcminute) because of the small-scale cutoff, but
with an amplitude that is even smaller than the decrease in the E-mode
signal. The parity-violating B-modes remain zero in the presence of a
small-scale cutoff.

These findings suggest that the aperture statistics introduce
only negligible E-/B-mode mixing for third-order shear statistics on the angular
scales that will be probed by ongoing and future weak lensing
surveys. Therefore, third-order aperture statistics can safely be
used as E-/B-mode separating statistics to infer cosmological information
from those surveys.

\begin{acknowledgements}
We thank Martin Kilbinger for helpful discussions, and Masahiro Takada for his
quick email response regarding their halo model paper \citep{TJ03c}. This work
was supported by the Deutsche Forschungsgemeinschaft under the project B5 in
the TRR33 `The Dark Universe'. BJ acknowledges support by an STFC Ernest
Rutherford Fellowship, grant reference ST/J004421/1. 

\end{acknowledgements}

\bibliographystyle{aa}
%\bibliography{bibliography}

\appendix
\section{Shape noise contribution to the uncertainty of the $\ba{M_{\rm ap}^3}$
measurement}
\label{sec:app}
Here we derive a simple formula that yields an approximate estimate of
$\sigma\br{\ba{M_{\rm ap}^3}}_{\rm s}$, the uncertainty of $\ba{M_{\rm ap}^3}$
measurement contributed by the shape noise.
 
We consider first the measurement of $\ba{M_{\rm ap}^3}$ within a single
aperture of angular radius $\theta$. A convenient estimator is
\eq{
\label{eq:Map3esti}
\hat{M}_{\rm ap}^3 = \br{\frac{1}{\bar{n}}}^3 \sum\limits_{i \ne j\ne k}
Q_{\theta, i} Q_{\theta, j} Q_{\theta, k}\; \epsilon_i \epsilon_j \epsilon_k \,,
}
where $\bar{n}$ is the mean number density of galaxy images, and the sum runs
over all triples of galaxy shapes in the aperture. This estimator resembles
Eg.\;(5.24) of \citet{s98}, but excludes the cases with two or three equal
indices, for example, $i=j$.
It is unbiased when the number of galaxies inside the aperture is
very large.

The shape noise contribution to the dispersion of $\ba{M_{\rm ap}^3}$
measurements in this aperture is, by definition, $\sigma\br{\ba{M_{\rm ap}^3}}_{\rm s, 1\rm f} =
\sqrt{\ba{\br{\hat{M}_{\rm ap}^3 }^2}_{\rm s}}$.
Neglecting high-order correlations, the shape noise contribution to
$\ba{\br{\hat{M}_{\rm ap}^3 }^2}$ is 
\eqs{
\label{eq:map6}
& \ba{\br{\hat{M}_{\rm ap}^3 }^2}_{\rm s} = \; \br{\frac{1}{\bar{n}}}^6 
 \br{\frac{\sigma_{\epsilon}^2}{2}}^3 \\
 & \times\; \ba{ \sum\limits_{i,j,k} \sum\limits_{l,m,n} 
Q_{\theta, i} Q_{\theta, j} Q_{\theta, k}  Q_{\theta, l} Q_{\theta, m}
Q_{\theta, n}\; \br{\delta_{il}\delta_{jm}\delta_{kn} + 5\ \rm{perms}} } \\
& =  \; 6 \br{\frac{\sigma_{\epsilon}^2}{2}}^3 \br{\frac{1}{\bar{n}}}^6
\textsc{P}\br{\sum\limits_{i=1}^{N_{\rm g}} Q_{\theta, i}^2 }^3 \\
& =  \; 6 \br{ \frac{\sigma_{\epsilon}^2}{2 \bar{n}} \int\dd^2\vartheta\;
Q_{\theta}^2(\vartheta)  }^3\,, }
where $N_{\rm g} = \bar{n} \pi \theta^2$ is the average number of galaxies
inside the aperture. In the second equation we substituted the ensemble average
$\ba{\ }$ with an average over all possible galaxy positions \citep[see
equation\;5.3 of][]{s98}, 
\eq{ \textsc{P}\br{X} = \br{\prod\limits_{i=1}^{N_{\rm g}} \int
\frac{\dd^2\vartheta_i}{\pi \theta^2}} X \,.
}

With \citet{crittenden02} filter function $Q_{\theta}$
(\ref{eq:Qform}), the integral in the last line of (\ref{eq:map6}) has a value
of $1/(\pi\theta^2)$. This leads to 
\eq{
\sigma\br{\ba{M_{\rm ap}^3}}_{\rm s, 1\rm f} =
\sqrt{\frac{6}{\pi^3}}\frac{1}{\theta^3} \br{\frac{\sigma_{\epsilon}^2}{2
\bar{n}}}^{3/2}\,.
}

For a galaxy survey of size $A$, a number $N_{\rm f}$ of nearly independent
apertures can be put on the field. This approximately reduces the uncertainty of $\ba{M_{\rm
ap}^3}$ measurement by a factor of $1/\sqrt{N_{\rm f}}$. Using a rough
estimation of $N_{\rm f} = A/(4 \theta^2)$, we obtain
\eq{
\label{eq:sigmap3shot}
\sigma\br{\ba{M_{\rm ap}^3}}_{\rm s} \approx \frac{\sigma\br{\ba{M_{\rm
ap}^3}}_{\rm s, 1\rm f}}{\sqrt{N_{\rm f}}} = 
\sqrt{\frac{24}{\pi^3 A}}\frac{1}{\theta^2} \br{\frac{\sigma_{\epsilon}^2}{2 \bar{n}}}^{3/2} \,.
}

We note again that this estimation of $\sigma\br{\ba{M_{\rm ap}^3}}_{\rm s}$ is
aimed to be simple and approximate. Practically, $\ba{M_{\rm ap}^3}$ is
estimated not with (\ref{eq:Map3esti}) but is derived from the shear 3PCFs
(\ref{eq:map_3}), or similarly, from the bispectrum of shear. This would lead to
a slightly different prefactor in the expression of $\sigma\br{\ba{M_{\rm ap}^3}}_{\rm s}$. For a more precise
estimation of this prefactor, one can use Eq.\;(26) in \citet{joachimi09b} as an
estimate of the bispectrum covariance, and derive the covariance of $\ba{M_{\rm
ap}^3}$ with the help of Eq.\;(46) in \citet{s05}. Another derivation of
$\sigma\br{\ba{M_{\rm ap}^3}}_{\rm s}$ for aperture mass statistics computed
with the polynomial filter function is provided in the appendix of \cite{kilbinger05}.

\end{document}